\documentclass[
 reprint,
amsmath,amssymb,
 aps,
prb,
superscriptaddress,
showpacs, showkeys]{revtex4-1}

\usepackage{xr-hyper}

\makeatletter
\newcommand*{\addFileDependency}[1]{% argument=file name and extension
	\typeout{(#1)}
	\@addtofilelist{#1}
	\IfFileExists{#1}{}{\typeout{No file #1.}}
}
\makeatother

\usepackage{color}
\usepackage{graphicx}
\usepackage{dcolumn}
\usepackage{bm}
\usepackage{hyperref}
\usepackage[normalem]{ulem}
\hypersetup{bookmarksnumbered, pdfpagemode=UseOutlines, 
colorlinks=true, citecolor=blue, filecolor=blue, linkcolor=blue, urlcolor=blue}

\graphicspath{X:\paper
}

\begin{document} 

\title{Strain driven antiferromagnetic exchange interaction in SrMnO$_3$ probed by phase shifted Spin Hall magnetoresistance}

\author{J. J. L. van Rijn}
\thanks{corresponding author}
\email{j.j.l.van.rijn@rug.nl}
\affiliation{University of Groningen, Zernike Institute for Advanced Materials, 9747 AG Groningen, The Netherlands}
\author{D. Wang}
\affiliation{Department of Physics and Astronomy, Uppsala University, Box-516, 75120 Uppsala, Sweden}
\author{B. Sanyal}
\affiliation{Department of Physics and Astronomy, Uppsala University, Box-516, 75120 Uppsala, Sweden}
\author{T. Banerjee}
\thanks{corresponding author}
\email{t.banerjee@rug.nl}
\affiliation{University of Groningen, Zernike Institute for Advanced Materials, 9747 AG Groningen, The Netherlands}

\date{\today}

\begin{abstract}
Multiferroics have found renewed interest in topological magnetism and for logic-in-memory applications. Among them, SrMnO$_{3}$, possessing strong magnetoelectric coupling is gaining attention for the design of coexisting magnetic and polar orders upon straining. Here we demonstrate antiferromagnetic exchange interactions in strained SMO thin films extracted from a new feature in the phase response of Spin Hall magnetoresistance, which has not been explored in earlier works, such as in magnetic insulators. We explain our findings with a model that incorporates magnetic anisotropy along [110] direction, corroborates with DFT studies and is consistent with the direction of ferroelectric polarization in SrMnO$_{3}$. The fundamental insights obtained from our studies establishes the potential of this material in magnetoelectrically coupled devices for different logic and memory applications.

\end{abstract}


\maketitle

\section{Introduction}

\par The ability to design, stabilize and tune diverse order parameters in the same material system renders complex oxides useful for the study of emerging properties, relevant for alternative computing strategies \cite{Manipatruni2019, Spaldin2019, Vaz2022}. In this respect, multiferroic materials, due to their simultaneous coupling between charge and magnetic order, enables electric field control of magnetization and spin transport,  offering new prospects for low power memory, logic or logic in memory applications. Less investigated is the antiferromagnetic order, intrinsic to such multiferroics, primarily due to it being unresponsive to modest magnetic fields. These studies are important not only for developing antiferromagnetic spintronics and topological magnetism but also for unraveling the full potential of magnetoelectric coupling in these materials by stimulating different routes to tailor diverse magnetic order. This can be realized in multiferroic materials of complex oxides by modulating the strain, crystal structure and oxygen defects.\\  
The rare earth manganite SrMnO$_{3}$ (SMO) is a magnetic insulator and a G-type antiferromagnet 
with the magnetic and ferroelectric order linked to the same B-site Mn cation, resulting in a strong magnetoelectric coupling. From first principle studies, a temperature and strain dependent phase diagram was recently established for SMO, highlighting the tunability of the magnetoelectric coupling with strain \cite{Edstrom2020,Edstrom2018}. At strains above 1.6\%, a ferroelectric phase with a substantial polarization is predicted due to orbital reordering and cation displacement, persisting upto room temperature for higher strains\cite{Edstrom2020,Edstrom2018,Lee2010,Zhu2020}. At coinciding polar and magnetic ordering temperatures, SMO promises sizable magneto-electric coupling \cite{Edstrom2020} making it an appealing material for applications.\\

Experimentally, the polar character in strained SMO films has been studied recently using various charge mediated experimental techniques \cite{Guzman2016,Schaab2016,Becher2015,Guo2018,An2021,Wang2020}. However probing the antiferromagnetic order in similarly strained SMO thin films has proved to be challenging due to the coexistence of competing magnetic phases. This is compounded by the fact that fabrication of stoichiometric strained SMO thin films is non trivial due to the formation of oxygen vacancies which relaxes the strain \cite{Langenberg2021,Kaviani2022,Wang2016,Agrawal2016}. For oxygen deficient films, the Manganese 4$^{+}$ oxidation state is reduced by the absence of negative oxygen atoms, promoting double exchange interaction and a ferromagnetic order.\\ 

Here we show how a new feature in the phase response of the Spin Hall Magnetoresistance (SMR), hitherto unreported, in tensile strained SMO thin films is explained by considering a coexistence of different antiferromagnetic domains, governed by magnetocrystalline anisotropy in SMO and complemented by bulk magnetization studies. The magnetic order is probed by analyzing both the phase and amplitude of the longitudinal SMR oscillations. SMR studies on SMO, until now, are found to be consistent with a ferromagnetic order attributed to oxygen deficient SMO phases \cite{Das2021}.\\

To understand different magnetic interactions that coexist in our strained SMO thin films, Density Functional Theory (DFT) studies are performed both for stoichiometric and oxygen deficient SMO, revealing remarkably large bond angle modulations, close to the film surface. We develop a model that incorporates magnetic anisotropy along [110] direction, determined from the rotational dependence of the SMR, corroborated by DFT studies and aligns well with the recently predicted \cite{Edstrom2020,Edstrom2018} ferroelectric polarization along the same direction in SMO.\\

\section{Film growth and characterization}

\begin{figure}[h!]
    \centering
    \includegraphics[width=.98\linewidth]{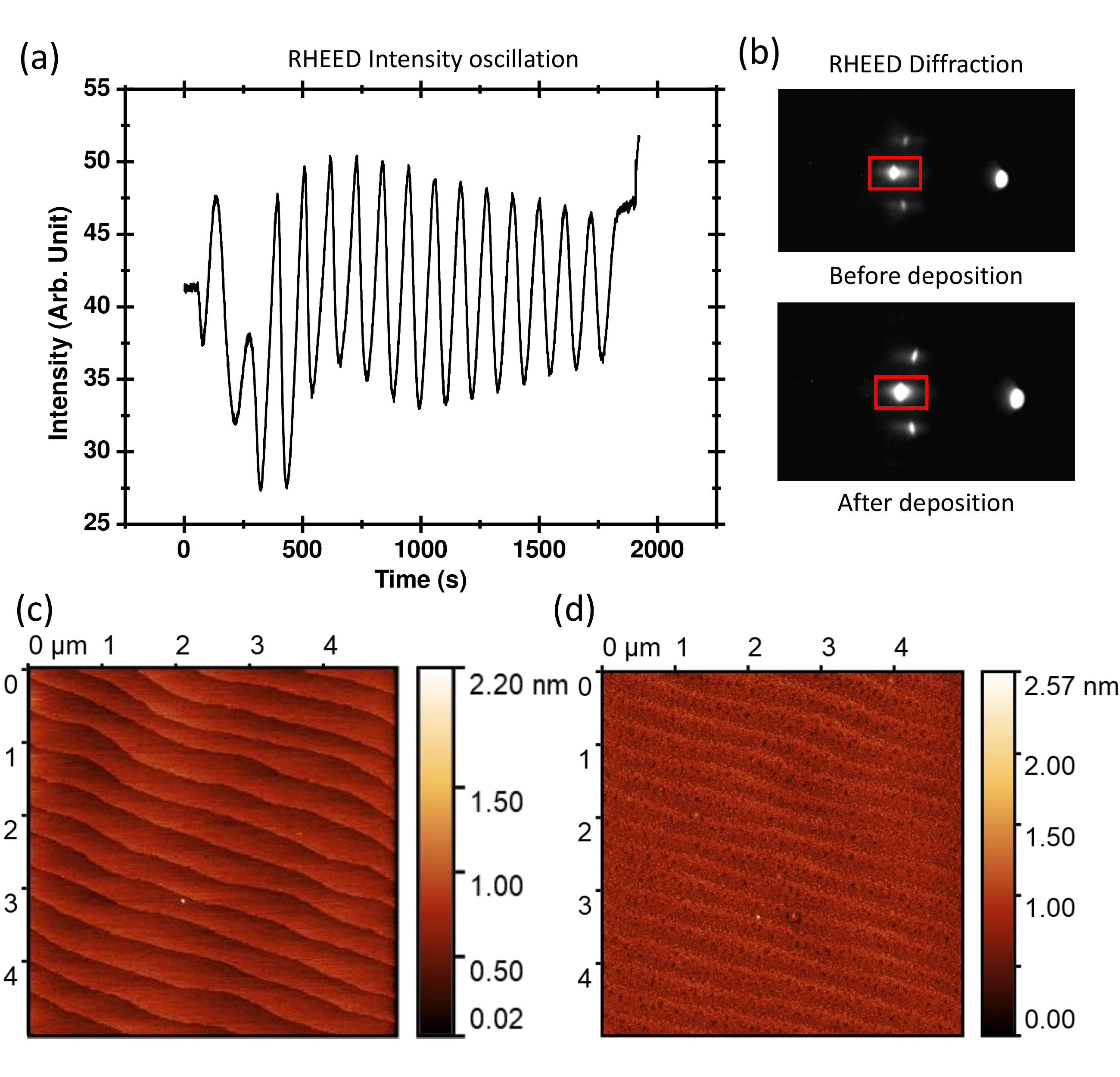}
	\caption{In a), in-situ Reflective High Energy Electron Diffraction intensity oscillations are displayed. The oscillations indicate clear layer-by-layer growth and from this a film thickness of 15 u.c. is determined. The diffraction spot used to obtain the intensity oscillations are shown in b), marked with a red square. The AFM images of a STO (001) substrate (c) and SMO film in (d) both display a low roughness which indicates the atomically flat surface. The STO terrace structure is clearly visible in (d).}
	\label{Fig_1}
\end{figure}

SrTiO$_{3}$ (001) substrates are prepared by a standard protocol to obtain a TiO$_{2}$ terminated surface\cite{Koster1998}. A topographic Atomic Force Microscopy (AFM) image of a terminated substrate is shown in figure \ref{Fig_1}c, indicating the terrace structure with a miscut of 0.05$^{\circ}$. The surface roughness, calculated over the full scan, has a Root Mean Sqaure (RMS) of 0.131 nm, indicating an atomically flat surface. The SrMnO$_3$ (SMO) thin films are grown by pulsed laser deposition at a substrate temperature of 800 $^{\circ}$C and an oxygen pressure of 0.05 mbar using a KrF laser with a laser fluence of 2 Jcm$^{-2}$. The films are post-annealed at 600 $^{\circ}$C for 30 minutes at an oxygen pressure of approximately 150 mbar. The surface structure of the SMO films is monitored using in-situ Reflective High Energy Electron Diffraction (RHEED) (figure \ref{Fig_1}). From the RHEED intensity oscillations we confirm layer-by-layer growth and ascertain the thickness of 15 u.c.. The SMO film surface roughness of 0.145 nm, see figure \ref{Fig_1}d, is comparable to the pristine substrate before growth, confirming high quality of the films.

\begin{figure}[h!]
    \centering
	\includegraphics[width=.98\linewidth]{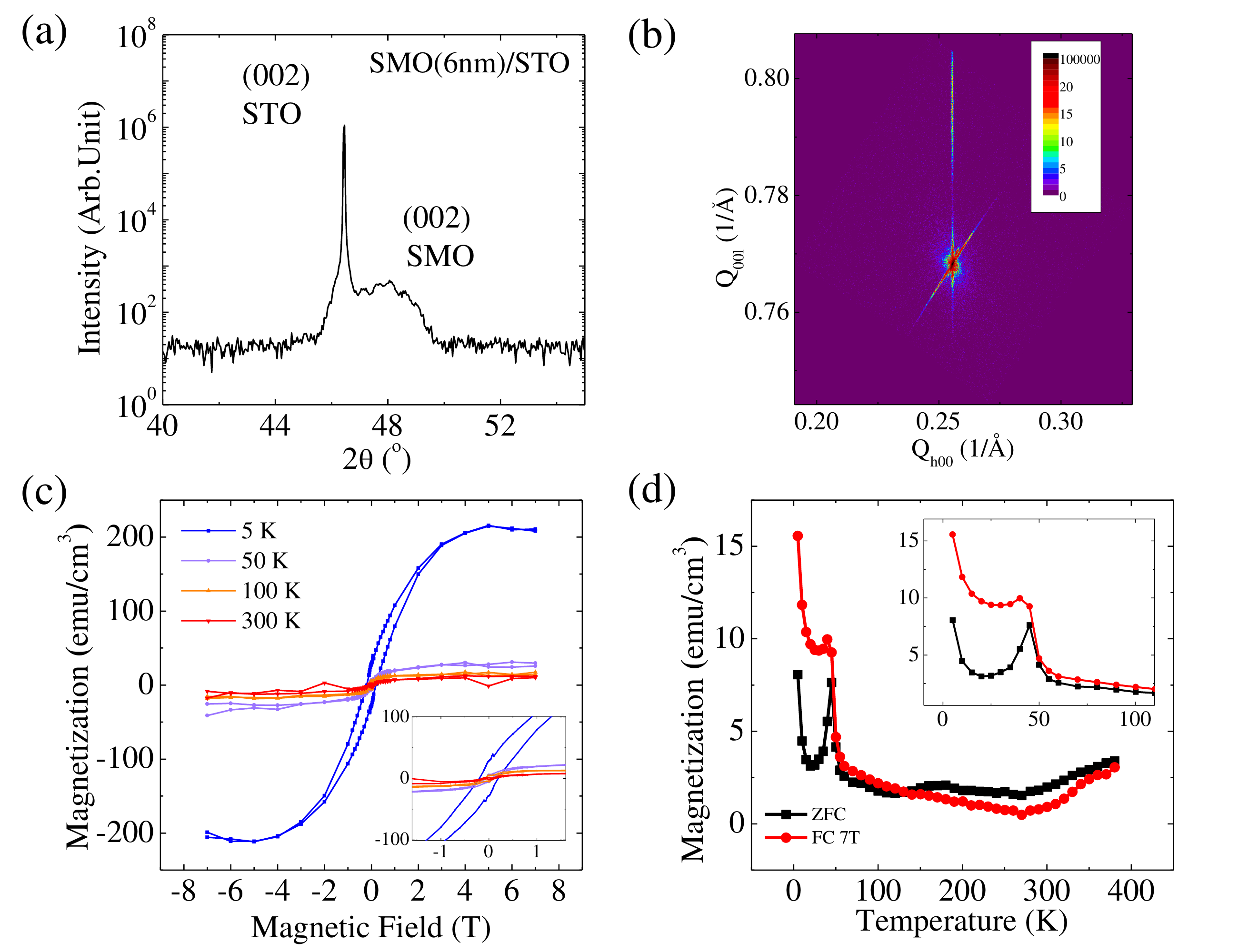}
	\caption{The (002) SMO peak in (a) obtained from X-Ray Diffraction corresponds to -0.6 \% strain. From the reciprocal space mapping in (b) a fully epitaxial strain of 2.6\% is confirmed. Magnetization versus temperature (M-T) for Zero Field Cooled (ZFC) and Field Cooled (FC) measurements is shown in (c). A measurement field of 1 kOe was applied during warming along the [100] crystallographic direction. In (d), the field dependent magnetization (M-H) complements the (M-T), displaying a ferromagnetic loop only for 5 K. In the insets a zoom of the respective measurement curve is shown.}
	\label{Fig_2}
\end{figure}

Using a four-axis cradle (Cu k${\alpha}$, ${\lambda}$ = 1.54) PANalytical X-ray diffractometer, 2${\theta}$ diffraction spectra are recorded at room temperature, shown in figure \ref{Fig_2}. From the 2${\theta}$ SMO peaks a fully strained perovskite SMO structure is confirmed, with an out-of-plane and in-plane strain of -0.6 \% and 2.6 \% respectively.

\par To understand the impact of the oxygen vacancies on the magnetic order in SMO, bulk magnetization measurements are utilized. Using a Quantum Design SQUID magnetometer, in-plane magnetization loops are obtained for temperatures ranging from 5 K to 300 K, displayed in figure \ref{Fig_2}c. At 5 K, a tiny loop opening in M-H is observed and interpreted to arise from vacancy driven weakening of the antiferromagnetic order. A net magnetization is observed between 5 K and 50 K, from the finite difference of magnetization measured for the Zero-Field-Cooled curve and Field-Cooled curve  shown in figure \ref{Fig_2}d. Above 50 K, both curves meet and the magnetization decreases further and disappears at higher temperature. 

\section{Computational details}

\par Our simulations are based on first-principles electronic structure calculations within density functional theory (DFT). We first performed geometry optimization for the structure with fixed bottom two layers by employing the projector augmented wave (PAW)\cite{PAW, PAW2} based density functional code VASP\cite{VASP1}. The plane-wave energy cutoff was set to 520 eV, and the force on each atom is converged to 0.01 eV/\r{A}. The k integration in the Brillouin zone was performed using $7\times7\times1$ points for geometry optimization and $12\times12\times1$ points for self-consistent calculations.
The optimized structure were then used as an input for the calculation of interatomic exchange parameters ($J_{ij}$) by means of the magnetic force theorem (MFT)\cite{MFT} using full-potential linear muffin-tin orbital (FP-LMTO) code RSPt\cite{rspt}. 
To describe the exchange-correlation effects, we used the generalized gradient approximation (GGA) in the form of Perdew, Burke, and Ernzrhof (PBE)\cite{pbe} augmented by the Hubbard-U corrections (PBE+U)\cite{DFT+U, DFT+U2}. The Coulomb U and intra-atomic Hund's exchange parameter J are added on Mn-\textit{d} electrons as 2.7 eV and 1 eV, respectively, which have proven to be  good values for Mn-based oxides\cite{lee2010epitaxial, zhu2020magnetic}.
Finally, we used the extracted $J_{ij}$ and magnetocrystalline anisotropy energies (MAEs) to calculate the magnetic ordering temperature using classical Monte Carlo (MC) simulations for the solution of the following spin Hamiltonian
$$
H=\sum_{i,j}J_{ij}e_i \cdot e_j - \sum_i K_i (\hat{e}_i\cdot e_i^{K})^2,
$$
which is implemented in the Uppsala Atomistic Spin Dynamics (UppASD) code\cite{uppasd}. For this purpose, a $70\times70\times1$ cell (58800 atoms) was considered. The transition state barriers are obtained by using the climbing image nudged elastic band (cNEB) method \cite{henkelman2000climbing, henkelman2000improved}. 
\par The structural model considered in this paper consists of six layers of SMO on top of three STO layers in the unit cell (figure \ref{Fig_3}a) with the vacuum region set to
20 \r{A}. Since the theoretically obtained equilibrium lattice parameters of bulk SMO and bulk STO are 3.945 \r{A} and 3.836 \r{A}, we fix the in-plane lattice parameter of the structural model as 3.945 \r{A}, that matches closely with the equilibrium lattice parameters of bulk SMO and STO and corresponds well with the experimental lattice mismatch of 2.9\%. In order to simulate structures with different magnetic configurations, such as, ferromagnetic (FM) and three antiferromagnetic (AFM), A-type, C-type, and G-type, a $\sqrt{2}\times\sqrt{2}$ in-plane supercell is chosen. 

\section{Results and discussion}

\begin{figure}[h!]
	\includegraphics[scale=1]{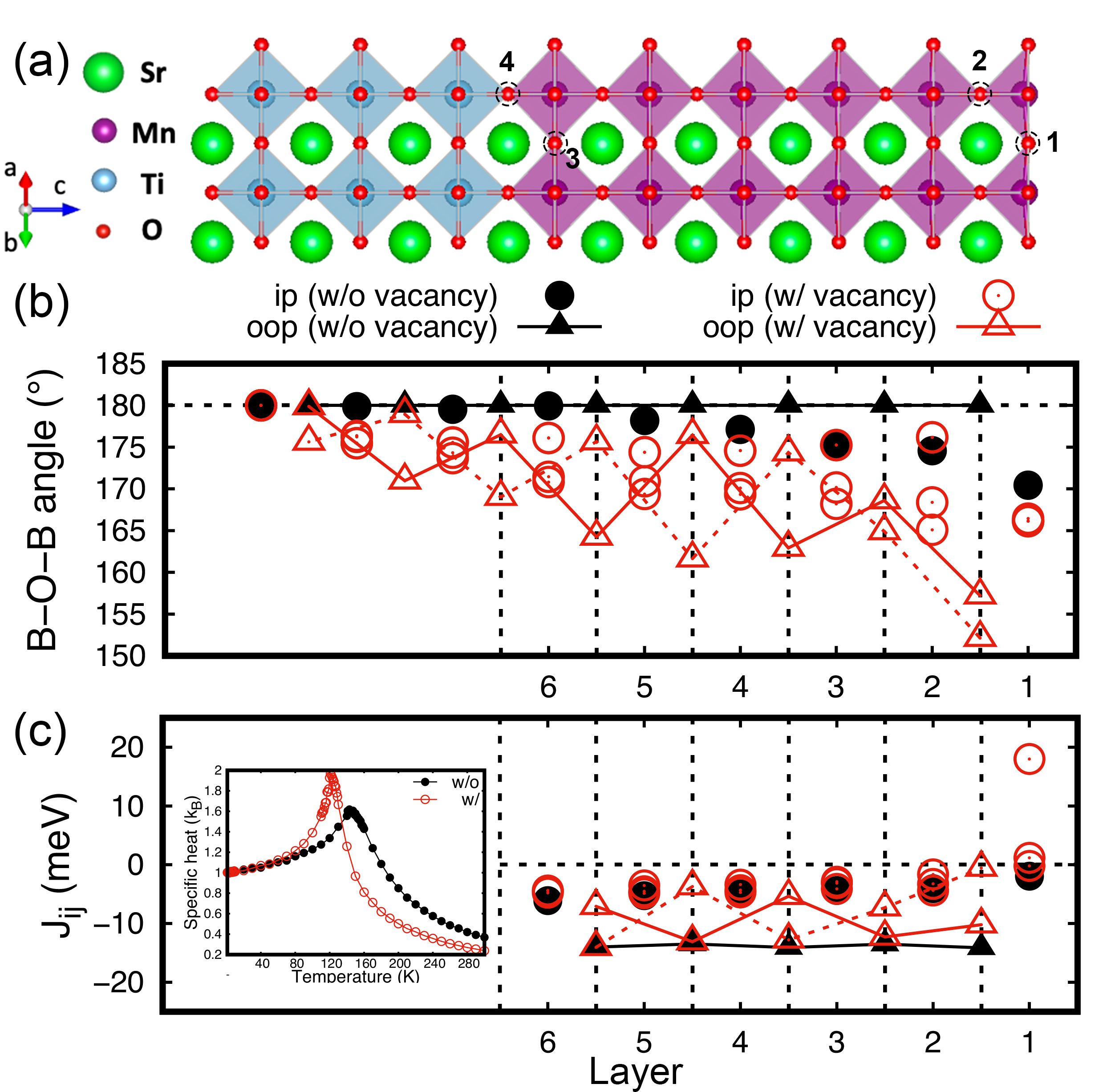}
    \caption{(a) The side view of SrMnO$_3$/SrTiO$_3$ used in DFT+U calculations. Green, purple, cyan, and red balls represent Sr, Mn, Ti, and O atoms, respectively. The shaded areas indicate TiO$_6$ and MnO$_6$ octahedra. Dashed circles indicate the positions of oxygen vacancy. (b) The Mn-O-Mn bond angles and (c) the nearest neighbor magnetic exchange parameters for structures with (filled circle and triangle) and without (empty circle and triangle) oxygen vacancy. Circle and triangle represent Mn-Mn pair in- (ip) and out-of-plane (oop) MnO$_2$ plane. In order to distinguish two different spin chains along the \textit{z}-direction, empty triangles are connected by solid and dashed lines. Vertical dashed lines are used to separate the MnO$_2$ layers. Positive and negative values of $J_{ij}$ in (c) indicate FM and AFM coupling, respectively. The inset in (c) is the specific heat as a function of temperature obtained from Monte Carlo simulations. In (b) and (c), the MnO$_2$ layers are indicated by numbers with the surface (interface with STO) marked by 1(6).}
    \label{Fig_3}
\end{figure}

Unlike bulk SMO, there are some interesting features for strained thin films on a substrate, linked to the Mn-O-Mn bond angles and are essential to the understanding of the exchange mechanism and the exchange parameters.
The in-plane Mn-O-Mn angle, as shown by filled circles in figure \ref{Fig_3}b, gradually decreases from 180$^\circ$ (the sixth layer) to 170$^\circ$ (the surface layer), which directly leads to the consequence in figure \ref{Fig_3}c, that the AFM coupling changes from -6.10 meV to -1.97 meV. This follows from the Goodenough-Kanamori-Anderson (GKA) rule that the AFM exchange is weakened by reducing the Mn-O-Mn angle from 180$^\circ$. On the other hand, all the out-of-plane angles keep the same value as the bulk SMO (180$^\circ$). Therefore, the exchange parameters remain almost unchanged (as shown in figure \ref{Fig_3}c with the black triangles). Further, we find the magnitude of the out-of-plane exchange parameter is, in general, larger than the in-plane exchange parameters. This is because the nearest Mn-Mn distances parallel and perpendicular to the interface are different due to the tensile strain introduced by the STO substrate. The averaged c/a ratio of the SMO thin film amounts to 0.96 from geometry optimization.

\par Due to the tensile strain introduced by the STO substrate, and depending on the growth conditions of the thin films, oxygen vacancies tend to be the dominant type of defect in this material. In our DFT simulations, we created oxygen vacancies at four different sites (see figure \ref{Fig_3}(a)) and performed geometry optimization for each structure. Results are shown in Table \ref{energies}. First, as shown in the first column (Site 1), regardless of the magnetic configuration, the oxygen vacancy tends to be at the surface layer. Second, the ground state structure is  G-type AFM with the oxygen vacancy at the surface layer. Third, the energy of the structure with oxygen at the MnO$_2$ layer (Site 1 and Site 3) is much smaller than the case on the TiO layer (Site 2 and Site 4). In order to estimate the energy involved in the diffusion of oxygen vacancy from one layer to another, we performed Nudged Elastic Band (NEB) calculations. They correspond to the vacancy transition from the top layer (site 1) to the layer below (site 2) and also from the six layer (site 3) to the interface (site 4). The transition barriers are 1.9 and 3.2 eV, respectively. It indicates that the vacancy is easy to move when close to the surface, while it is hard when present deep inside the thin film. Besides that, the most interesting result is the magnetic property. As it is shown in figure \ref{Fig_3}(c), instead of having AFM coupling, the surface layer which contains an oxygen vacancy shows an FM coupling. This result is observed in the self-consistent calculations and also confirmed by the calculated exchange parameters. Combined with the increased magnitude of the moment on one of the Mn atoms (from 2.60 to 3.77 $\mu_B$), the strength of this FM coupling is stronger than all the other couplings (18 meV compared to, for example, -4.35 meV from the second MnO$_2$ layer). 
\par In addition to the changes in crystal structure and magnetic exchange parameters described in the main paper, the presence of an oxygen vacancy also affects the ground-state magnetic structure as well as the magnetocrystalline anisotropy energy of SMO. Results are shown in Table \ref{energies} and \ref{table_mae}.

\begin{table}[htb]
\caption{\label{energies}Calculated relative energies of structures with oxygen vacancy. The ground state energy is set to zero. Numbers in parenthesis are the oxygen vacancy formation energies. They are calculated by E$_{form}$=E$_{w/}$-E$_{w/o}$-E$_{oxygen}$, unit in eV.}
\begin{ruledtabular}
\begin{tabular}{ccccc}
\textrm{ }&
\textrm{Site 1}&
\textrm{Site 2}&
\textrm{Site 3}&
\textrm{Site 4}\\
\colrule
A-type & 0.19(0.01) & 0.74(0.56) & 1.14(0.96) & 2.09(1.91) \\
C-type & 0.19(0.04) & 0.73(0.58) & 1.62(1.47) & 2.33(2.18) \\
G-type & 0.00(0.02) & 1.13(1.15) & 1.54(1.56) & 2.43(2.45) \\
FM     & 0.16(-0.41) & 1.07(0.50) & 1.09(0.52) & 2.42(1.86) \\
\end{tabular}
\end{ruledtabular}
\end{table}

\begin{table}[hptb]
\caption{\label{table_mae}Calculated magnetocrystalline anisotropy energy for the SMO structures with and without oxygen vacancy.
The MAE is computed by $E_{MAE}= E_{hard} - E_{easy}$, and the energy for the easy axis is set to zero. Unit in meV.}
\begin{ruledtabular}
\begin{tabular}{cccccc}
\textrm{ }&
\textrm{[100]}&
\textrm{[010]}&
\textrm{[110]}&
\textrm{[$\bar{1}$10]}&
\textrm{[001]}\\
\colrule
w/o vacancy & 0.3 & 0.3 & 0.0 & 0.0 & 475.8 \\
w/ vacancy& 221.6 & 221.4 & 0.0 & 1371.6 & 84.0 \\
\end{tabular}
\end{ruledtabular}
\end{table}

\par Furthermore, we found that the vacancy introduces significant distortions in the crystal structure. These distortions are evident in the changes to the B-O-B angles (see figure \ref{Fig_3}(b)). Both in-plane (empty circles) and out-of-plane angles (empty triangles) have drastically different values compared to the structure without vacancy (filled circles and triangles). For the out-of-plane Mn-Mn pairs, there are two Mn chains along the z-direction. These two chains perform differently in the structure with oxygen vacancy: they oscillate alternately, one after another. This interesting phenomenon is reflected in the out-of-plane exchange parameters. For the in-plane Mn-Mn pairs, the symmetry of the structure is reduced due to the distortion. Therefore, the equivalent Mn-O-Mn angles in the structure without vacancy no longer exist. 
The behavior of corresponding $J_{ij}$s can be explained by the GKA rule: the bigger the deviation angle from 180$^\circ$, the weaker is the AFM coupling. 

\begin{figure*}
	\includegraphics[scale=0.65]{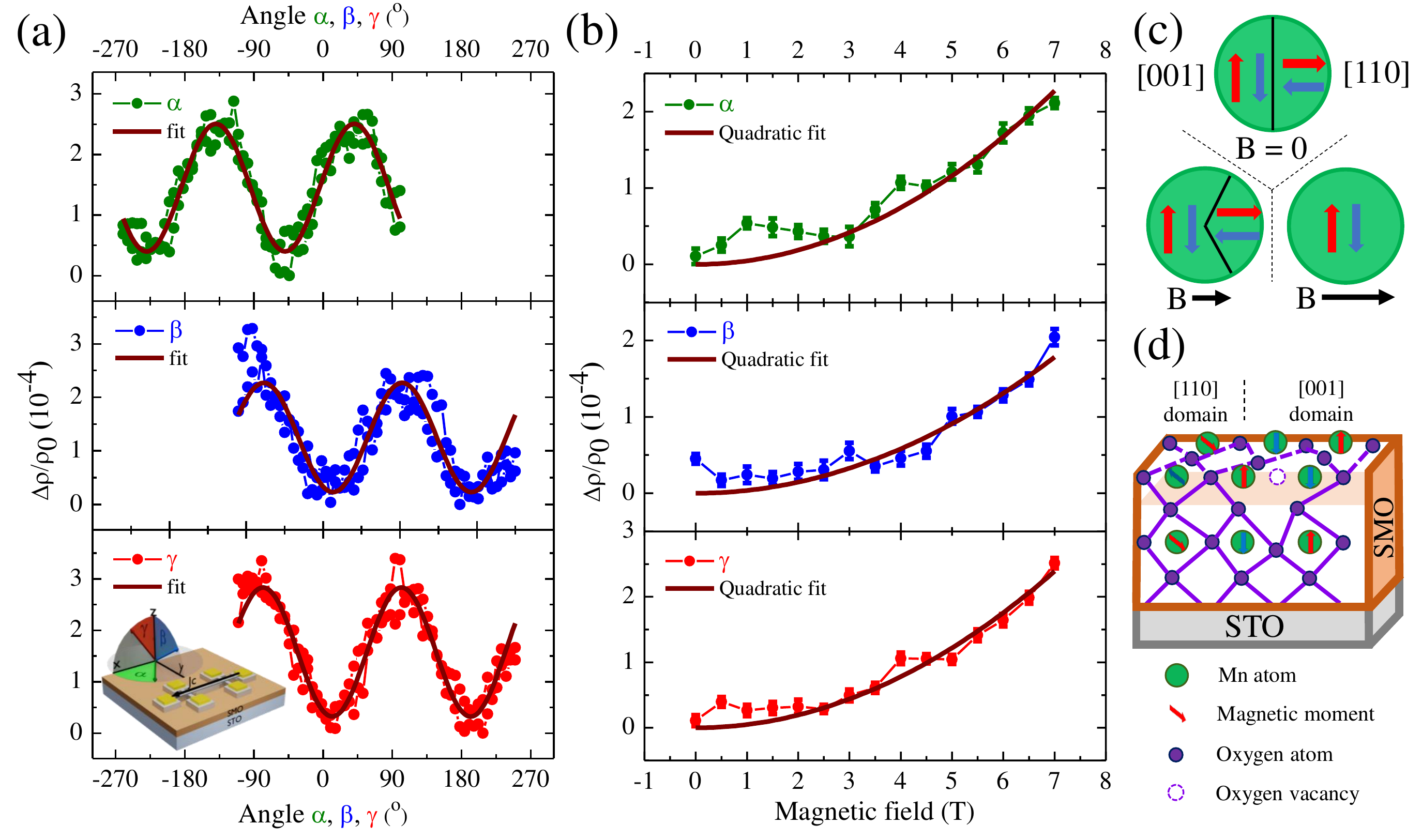}
	\caption{In the top panel of (a), a 135 degree phase shifted SMR curve is displayed for the $\alpha$ rotation, taken at 5 K using an external magnetic field of 7 T. The SMR in $\beta$ and $\gamma$ in the middle and bottom panels respectively both display a low resistance for out-of-plane angles. The curves are fitted with a $A*sin^2(x$-$x_{0})+y_{0}$ fit. In the $\gamma$ panel inset, a schematic illustration of the sample and Pt Hall bar device including definition of rotational directions is shown. In (b) the magnitude of the field dependent ADMR is shown for 5 K, which is fitted with a quadratic fit forced through the origin. The extracted quadratic coefficients for $\alpha$, $\beta$ and $\gamma$ are 4.6E-6 ± 2.6E-7, 3.6E-6 ± 2.3E-7 and 4.9E-6 ± 2.0E-7 respectively. Panel (c) depicts the redistribution of antiferromagnetic domains by increasing the strength of an externally applied magnetic field. In (d), a sketch of the magnetic order in SMO thin films is shown, incorporating the domain types with anisotropy directions.}
	\label{Fig_4}
\end{figure*}

\par To detect the magnetic ordering in strained SMO thin films, Spin Hall Magnetoresistance measurements are performed. SMR is an established method to probe surface magnetic order and for extracting useful parameters such as magnetic anisotropy and spin mixing conductance in ferromagnetic \cite{Vlietstra2013a,Velez2016}, antiferromagnetic \cite{Hoogeboom2017,Lebrun2019} and chiral magnetic insulators \cite{Aqeel2021,Aqeel2015}. In SMR, Spin Hall Effect (SHE) mediated spin accumulation at the interface of a heavy metal, typically Platinum (Pt), and an insulating magnetic material is reflected or absorbed based on the orientation of surface magnetic moments. Upon applying a rotating external magnetic field, the magnetic order in SMO is manipulated resulting in Angular Dependent Magnetoresistance (ADMR) in Pt \cite{Vlietstra2013a,uchida2010spin}. In this work, 8 nm thick Pt nanostructures (0.5x7 ${\mu}$m) are fabricated on the SMO thin films using electron beam lithography and DC sputtering. The temperature dependent Pt resistivity is shown in figure S4\cite{SI}. An ac charge current $J_c$ of 300 ${\mu}$A is applied in $\vec{x}$, corresponding to the [100] crystallographic direction as shown in figure \ref{Fig_4}a (inset, bottom panel). The generated spin accumulation in $\vec{y}$ is reflected or absorbed based on the orientation of the underlying magnetic moments resulting in a charge voltage via the Inverse Spin Hall Effect. This is measured as a resistance modulation in Pt, for antiferromagnetic materials governed by its N\'eel vector $\vec{l}$ as\cite{Chen2013,Fischer2018,Geprags2020}.

\begin{equation}\label{eq_AFneel}
    \rho_{long} = \rho_{0} + \frac{1}{2}\rho_{1}[1 - l_y^2] ,
\end{equation}

where $\rho_0$ is the intrinsic Pt resistivity, $\rho_1$ a SMR coefficient based on SHE related parameters such as the Spin Hall Angle and spin mixing conductance \cite{Vlietstra2013a}. The N\'eel vector is given by the difference of sublattice magnetic moments {m} as $\vec{l}$ = $(\vec{m}_1 - \vec{m}_2)/2$.
For applied fields strong enough to overcome the magnetic anisotropy, $\vec{l}$ orders perpendicular to the applied field direction allowing canting towards the magnetic field to decrease Zeeman energy. This results in a negative SMR $(sin^2(\alpha))$ for AFMs with respect to positive SMR ($-sin^2(\alpha$)) in FMs \cite{Hoogeboom2017,Lebrun2019} and with a phase shift of $90^{\circ}$. SMR measurements are performed for applied fields between 0 and 7 T at 5 K and for temperatures between 5 K and 300 K at a constant field of 7 T. The obtained ADMR curves, shown in figure \ref{Fig_4}a, are fitted with a sinusoidal dependence ($Asin^{2}(x-x_0)$) to extract the SMR magnitude ($A$) and phase ($x_{0}$). Remarkably at 5 K, a phase shift of 135$^{\circ}$ is observed in the ${\alpha}$ direction that cannot be explained by both AFM and FM SMR as shown in figure \ref{Fig_4}a. The obtained ADMR response in $\beta$ and $\gamma$ directions are strikingly similar and are not consistent with SMR for either ferromagnets or antiferromagnets, soft enough, for the magnetic order vector to follow the applied magnetic field. However, considerations of a non-trivial preferred direction of the magnetic anisotropy can explain our findings.  
\par By performing fully-relativistic simulations, we find the magnetic easy axis to be along [110] direction for both structures. The results are shown in Table S2. For the structure without oxygen vacancy, [-110] is an easy axis by symmetry of the tetragonal structure, the [001] direction is a hard axis with an anisotropy energy of 475.8 meV with the two in-plane orientations, [100] and [010] being equivalent with an anisotropy energy of 0.3 meV.
Interestingly, our results for the structure with oxygen vacancy show that the [100] energy is increased to an energy of 222 meV, while the out-of-plane energy is reduced significantly to an energy of 84 meV. Due to the breaking of inversion symmetry, [-110] becomes a hard axis with an energy of 1372 meV such that [110] and [001] become the easiest anisotropy axes.
\par We assess other contributions such as those arising from Hanle Magnetoresistance (HMR) and Weak Anti-localization (WAL) to our data. 

A sizable contribution from HMR at low temperature is excluded as this would result in a a 90$^{\circ}$ phase shift in  $\beta$ which is not the case. WAL though known to contribute in $\beta$ and $\gamma$ directions in Pt \cite{Velez2016,Phanindra2022} due to the strong spin-orbit coupling at temperatures below 50 K, is not a dominant contributor in our case as the out-of-plane direction has a low resistance with respect to the in-plane direction that shows a higher resistance. 
\par The quadratic dependence of SMR is unusual and compels us to consider the spatial orientation and distribution of antiferromagnetic domains in SMO with magnetic field.

Here we propose a model, where we consider that the spatial distribution of antiferromagnetic domains along [110] and [001] are manipulated by the external magnetic field. The choice for the anisotropy directions follows from the DFT calculations that predict a magnetic easy axis in the [110] direction and relatively low anisotropy energy in the [001] direction for oxygen deficient SMO.  Due to magnetic dipolar field energy, domains are expected to form in all magnetic materials, which affects the orientation of magnetic moments. Incorporating a domain fraction in equation \ref{eq_AFneel} gives: \cite{Fischer2018}

\begin{equation}\label{eq_SMRdomain}
    \rho_{long} = \rho_{0} +  \rho_{1}\sum_{k}\xi_k[1 - (l^k_y)^2]  
\end{equation}

where ${\xi}_{k}$ is the fraction of the magnetic domain k. The sum of the domain fractions should equal 1. A general expression of the fraction of the possible domains is given as:

\begin{equation}\label{eq_DomainFraction}
    \xi_k = \xi_0 \left[1 + \frac{2H^2}{H^2_{MD}}f(l_k,\delta)\right]
\end{equation} 

Here \textit{H} is the applied magnetic field and \textit{H$_{MD}$} is the magnetic field required to set a monodomain state in the magnetic volume\cite{Fischer2018}. Assuming both domain types are energetically degenerate, each occupies an equal volume in the absence of an applied field, described with \textit{$\xi_0$ = 1 / number of domains}. The function $f(l_k,\delta)$ is dependent on the relative direction between an applied magnetic field and the N\'eel vector of the particular domain, see figure S3c\cite{SI}. It governs the modulation of the resistance with the applied magnetic field such that the fraction of a domain is promoted when the N\'eel vector is perpendicular to the applied magnetic field. The field-induced Zeeman energy is then reduced by canting of the magnetic moments towards the applied magnetic field. A parallel alignment between the field and the Neel vector, on the other hand, is unfavourable resulting in a reduction of these domains. Consequently, an applied magnetic field will change the domain fraction within the fixed magnetic volume.
\par Two possible domains are considered, where the N\'eel vector is in-plane, parallel to [110] (k=1) and out-of-plane, parallel to [001] (k=2) as indicated in figure S3\cite{SI}

The SMR variation is obtained from the domain fractions ($\xi_k$) and the N\'eel vector projection onto $\vec{y}$ ($l^k_y$) for both domain types. This model assumes magnetic anisotropy is strong enough to have negligible spin canting such that when the field is applied perpendicular to the $(\bar110)$ plane, the domain fractions are not affected. Hence, the effective strength of the applied magnetic field is described by its projection onto the ($\bar110$) plane for the ${\alpha}$, ${\beta}$ and ${\gamma}$ rotations \cite{Geprags2020}. The resulting domain fractions and projected N\'eel vector are calculated as:

\begin{align}\label{eq_domaintype1}
    \xi_1 = a\left[1+\frac{2H^2}{H_{MD}^2}cos(2\delta)\right] && l^1_y = \frac{1}{\sqrt{2}}
\end{align}
and
\begin{align}\label{eq_domaintype2}
    \xi_2 = (1-a)\left[1+\frac{2H^2}{H_{MD}^2}cos(2\delta-\pi)\right]  && l^2_y = 0 ,
\end{align}

where \textit{a} (\textit{$0\leq a\leq1$}) is the fraction of ${\xi}_{1}$ without an applied field. $\delta$ is defined as the angle between the [001] direction and the projection of the magnetic field onto the ($\bar110$) plane, see figure S3c\cite{SI}. First we consider the rotation of the magnetic field in the $\alpha$ direction. By substituting the magnitude of the field, $|H_{(\bar110)}(\alpha)|$, and the angle between the applied field vector and the $(\bar110)$ plane in equations \ref{eq_domaintype1} and \ref{eq_domaintype2} and consequently in equation \ref{eq_SMRdomain}, we obtain the resistivity dependence for rotations in $\alpha$ as:

\begin{widetext}
\begin{align}\label{eq_SMRalpha}
    \rho_{long}(\alpha) = \rho_0 + \rho_{1}\left(\left(1-\frac{1}{2}a\right)+ 
    \left(1-\frac{3}{2}a\right)\frac{2H^2}{H_{MD}^2}cos^2\left(\frac{1}{4}\pi-\alpha\right)\right) .
\end{align}
\end{widetext}
Similarly, for rotations in $\beta$, the magnetic field magnitude and projection substituted yield for the domain fractions:

\begin{align}\label{eq_DFs_beta1}
    \xi_{1}(\beta) = a\left[1+\frac{2H^2}{H_{MD}^2}\left(\cos^2(\beta)-\frac{1}{2}\sin^2(\beta)\right)\right] 
\end{align}
\begin{align}\label{eq_DFs_beta2}   
    \xi_{2}(\beta) = (1-a)\left[1+\frac{2H^2}{H_{MD}^2}\left(-\cos^2(\beta)+\frac{1}{2}\sin^2(\beta)\right)\right].
\end{align}
By symmetry, rotations in $\beta$ and $\gamma$ are equal and hence the resistivity for both can be written as:

\begin{widetext}
\begin{align}\label{eq_SMRbetagamma}
    \rho_{long}(\beta, \gamma) = \rho_0 +  \rho_{1}\left(\left(1-\frac{1}{2}a\right)+\left(\frac{3}{2}a-1\right)\frac{2H^2}{H^2_{MD}}+\left(\frac{3}{2}-\frac{9}{4}a\right)\frac{2H^2}{H_{MD}^2}cos^2(\beta, \gamma)\right) .
\end{align}
\end{widetext}
To determine \textit{a}, the parameter $H_{MD}$ has to be known in addition to the SMR magnitudes of $\alpha$ and one out-of-plane rotational variation due to the interdependence of the rotational directions. However, the magnitudes of the oscillations are possibly impacted by other effects besides SMR such as WAL and HMR. Given that the $H_{MD}$ is not reached using an applied field of 7 T as no saturation of the SMR amplitude is observed, \textit{a} is not determined. Therefore we assume that both contribute equally (\textit{a} = 1/2). This reduces equations \ref{eq_SMRalpha} and \ref{eq_SMRbetagamma} to
\begin{align}\label{eq_SMRaishalfalpha}
    \rho_{long}(\alpha) = \rho_0 + \frac{3\rho_{1}}{4}\left(1+\frac{2H^2}{3H_{MD}^2}cos^2\left(\frac{1}{4}\pi-\alpha\right)\right) 
\end{align}
and
\begin{align}\label{eq_SMRaishalfbetagamma}
    \rho_{long}(\beta,\gamma) = \rho_0 + \frac{3\rho_{1}}{4}\left(1-\frac{2H^2}{3H^2_{MD}}+\frac{H^2}{H_{MD}^2}sin^2(\beta,\gamma)\right)
\end{align}
respectively. 
\par The resulting phase obtained from the model for the three rotational variations matches the phase as measured by ADMR. As also evident from \ref{eq_SMRaishalfalpha} and \ref{eq_SMRaishalfbetagamma} the magnitude of the SMR in  ${\alpha}$ should be smaller than for ${\beta}$ and ${\gamma}$, which is not the case in experiments, see figure \ref{Fig_4}b. This can occur due to either a difference between the anisotropy energies along [110] and [001] or non-negligible contributions from HMR and WAL.

\begin{figure}[]
	\includegraphics[width=.98\linewidth]{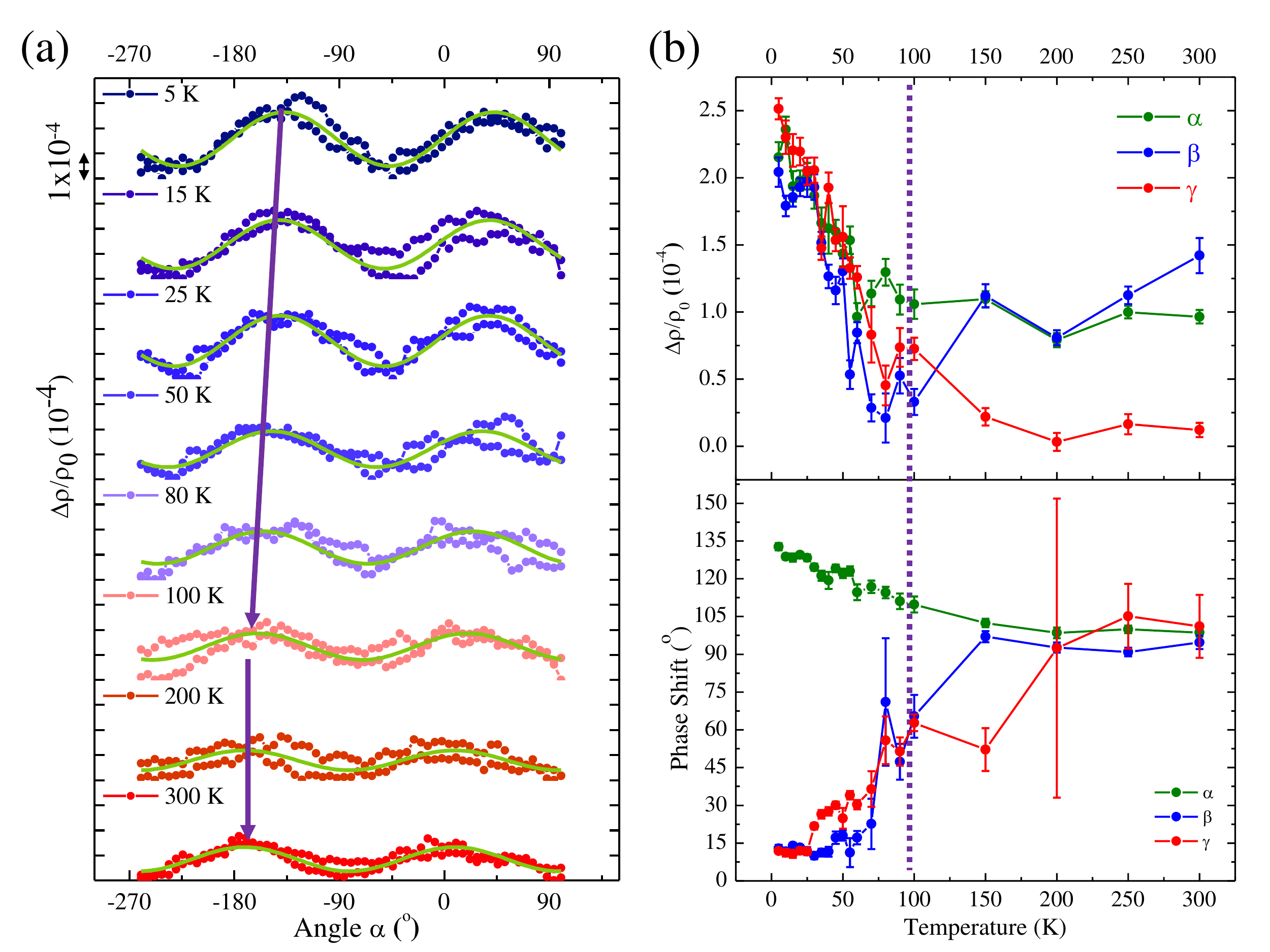}
	\caption{The temperature dependent SMR at 7 T for $\alpha$ in (a) indicates a gradual decrease of the phase shift up to 100 K as indicated with the purple arrows. The phase shift and magnitude is extracted by fitting the resistance modulation with  ${A sin^{2}(\alpha- x_0)}$ (green line). In (b) the extracted temperature dependent SMR magnitude and phase shift are displayed in the top and bottom panel respectively for all rotations. The error bars represent the fit error. Since there is no sizable signal for the $\gamma$ phase at 200 K, a remarkable large fit error is obtained.}
	\label{Fig_5}
\end{figure} 

\par Interestingly, we find a gradual decrease of the phase shift in SMR as the temperature increases from 5 K to 300 K (left panel of figure \ref{Fig_5}). In the right panel, the magnitude and phase extracted from the sinusoidal fits shows two distinct temperature regimes, indicated by the dashed lines at 100 K. Beyond 150 K, the finite magnitude and phase of 90$^{\circ}$ of the SMR signal in ${\alpha}$ and ${\beta}$, in addition to the diminishing SMR signal in  ${\gamma}$ indicates that the magnetic ordering in SMO is lost.
The disappearance of the SMR between 100 K and 150 K is in fair agreement with the calculated SMO ordering temperatures from the DFT-based simulations (as shown in the inset of figure \ref{Fig_3}c, the magnetic transition temperatures for stoichiometric and oxygen deficient SMO are 143 and 121 K, respectively). The gradual phase shift observed in all rotational directions is remarkable and suggests a modulation of the effective magnetic anisotropy with increasing temperature.
Suppression of specific anisotropy axes may arise due to possible temperature dependent structural properties such as polar order or strain similar to that reported in strained manganites as La$_{0.67}$Sr${_0.67}$MnO$_{3}$ where a strong temperature dependence of the magnetocrystalline anisotropy is observed \cite{Burema2021,Burema2022}. 

\section{Conclusions}

\par In conclusion, the antiferromagnetic character of strained SMO is revealed by phase shifted Spin Hall Magnetoresistance in three rotational directions and explained by a model based on magnetic anisotropy and spatial redistribution of antiferromagnetic domains with the external field. Temperature dependent SMR phase and magnitude affirms the gradual decrease of the magnetic anisotropy upon increasing temperature. The magnetic anisotropy in [110] and [001] directions, corroborated by DFT calculations is consistent with the ferroelectric polarization direction predicted for strained SMO. A detailed analysis of interface and surface induced structural distortions and corresponding modifications of exchange interactions obtained from DFT+U calculations has been provided. Also, it has been shown that an O vacancy at the surface of the film introduces a ferromagnetic order at the surface layer. The results obtained from this work constitute an important step towards a comprehensive understanding of the magnetic order in strained SMO films, which is crucial for the development of antiferromagnetic spintronics, orbitronics as well as for electric field control of magnetic order in multiferroic materials for alternative computing applications.

J. J. L. vR acknowledges financial support from Dieptestrategie grant (2019),  Zernike Institute for Advanced Materials. J. J. L. vR and T.B. acknowledges technical support from J. Baas, J. G. Holstein and H. H. de Vries and acknowledges A. Das, S. Chen, A. S. Goossens, M. A. Frantiu, G. E. W. Bauer and B. J. van Wees for useful discussions. This work was realized using NanoLab-NL facilities. The computations were enabled in project SNIC 2021/3-38 by resources provided to B.S. by the Swedish National Infrastructure for Computing (SNIC) at NSC, PDC and HPC2N partially funded by the Swedish Research Council through grant agreement no. 2018-05973. B.S. also acknowledges allocation for supercomputing hours by PRACE DECI-17 project Q2Dtopomat.
D.W. acknowledges the China scholarship council for financial support (No. 201706210084).

\bibliography{Bibliography}
\end{document}